\documentclass{aa}  
\usepackage{graphicx}
\usepackage{txfonts}
\usepackage{subfigure}
\usepackage[figuresright]{rotating}
\usepackage{textcomp}
\usepackage{txfonts}
\def\degr{\hbox{$^\circ$}}
\def\mnras{MNRAS}
\def\apj{ApJ}
\def\aap{A\&A}
\def\aaps{A\&AS}
\def\jrasc{JRASC}
\def\apjs{ApJS}
\def\aj{AJ}
\def\pasp{PASP}

\usepackage{natbib} 
\bibpunct{(}{)}{;}{a}{}{,}

\begin{document}

\title{High resolution optical spectroscopy of Plaskett's star\thanks{Based on observations made at the European Southern Observatory (La Silla, Chile) and at the Observatoire de Haute Provence (France).}}
\author{N. Linder\inst{1} \and G. Rauw\inst{1}\fnmsep\thanks{Research Associate FNRS, Belgium} \and F. Martins\inst{2} \and H. Sana \inst{3}\and M. De Becker\inst{1}\fnmsep\thanks{Postdoctoral Researcher FNRS, Belgium} \and E. Gosset\inst{1}\fnmsep$^{\star\star}$}

\offprints{N. Linder}

\institute{Institut d'Astrophysique et de G\'eophysique, Universit\'e de Li\`ege, B\^at. B5c, All\'ee du 6 Ao\^ut 17, B-4000 Li\`ege, Belgium\\
           \email{linder@astro.ulg.ac.be}
           \and 
           GRAAL, Universit\'e Montpellier II, CNRS, Place Eug\`ene Bataillon, F-34095 Montpellier, France
           \and
           European Southern Observatory, Alonso de Cordova 3107, Vitacura, Santiago 19, Chile\\
          }
\date{}

\abstract{Plaskett's star (\object{HD~47\,129}) is a very massive O~+~O binary that belongs to the Mon OB2 association. Previous work suggested that this system displays the Struve-Sahade effect although the measurements of the secondary radial velocities are very difficult and give controversial results. Both components have powerful stellar winds that collide, and produce a strong X-ray emission.}
{Our aim is to study in detail the physical parameters of this system, and to investigate the relation between its spectral properties and its evolutionary status.}
{We present here the analysis of an extensive set of high resolution optical spectra of \object{HD~47\,129}. We use a disentangling method to separate the individual spectra of each star. We derive a new orbital solution and discuss the spectral classification of both components. A Doppler tomography technique applied to the emission lines H$\alpha$ and \ion{He}{ii}~$\lambda$~4686 yields a Doppler map that illustrates the wind interactions in the system. Finally, an atmosphere code is used to determine the different chemical abundances of the system components and the wind parameters.}
{\object{HD~47\,129} appears to be an O8~III/I~+~O7.5~III binary system in a post RLOF evolutionary stage, where matter has been transferred from the primary to the secondary star. The He overabundance of the secondary supports this scenario. In addition, the N overabundance and C underabundance of the primary component confirm previous results based on X-ray spectroscopy and indicate that the primary is an evolved massive star. We also determine a new orbital solution, with $M_{\mathrm{P}}\,\sin^3i = 45.4 \pm 2.4$~M$_{\odot}$ and $M_{\mathrm{S}}\,\sin^3i = 47.3 \pm 0.3$~M$_{\odot}$. Furthermore, the secondary star has a large rotational velocity ($v\,\sin\,i \sim 300$~km~s$^{-1}$) that deforms its surface, leading to a non-uniform distribution in effective temperature. This could explain the variations in the equivalent widths of the secondary lines with phase. We suggest that the wind of the secondary star is confined near the equatorial plane because of its high rotational velocity, affecting the ram pressure equilibrium in the wind interaction zone.}{}

\keywords{stars: individual: HD\,47129 -- binaries: spectroscopic -- stars: fundamental parameters}

\maketitle

\section{Introduction}

Plaskett's star (\object{HD~47\,129}) is a non-eclipsing early type binary which was analysed for the first time by \citet{Plaskett22}. He found doubled spectral lines and described the system as consisting of a bright O5e primary component and a fainter secondary component with weak and diffuse lines. When he determined the elements of the spectroscopic orbit, the large mass of this system attracted his attention ($(M_{\mathrm{P}} + M_{\mathrm{S}}) \sin^3 i = 138.9$~M$_{\odot}$). This result motivated \citet{Struve48} to re-observe this object in 1947$-$1948 over two complete orbital cycles. He detected that some lines in the secondary spectrum (H$\gamma$, \ion{He}{i}~$\lambda\lambda$~4026,~4471) were weaker when this component was receding. This effect is nowadays called the Struve-Sahade effect (S-S effect hereafter) and is currently defined as the apparent strengthening of the secondary spectrum of a binary system when the secondary is approaching, and the corresponding weakening of its lines when it is receding. \citet{Struve48} also noted differences in the intensity of the secondary spectrum from one cycle to another, at a given orbital phase. He explained these variations with the presence of streams of gas in the system. A third observational campaign, conducted between 1956 and 1958 by \citet{SSH58}, allowed these authors to conclude that the behavior of the radial velocities of the secondary component suggests that it ejects matter in an irregular manner, and that these ejections are larger over the hemisphere facing the primary star. \citet{HC76} confirmed the presence of the S-S effect in the system and suggested that the primary star has a stellar wind which is stronger on the rear side of the star with respect to its orbital motion. IUE data allowed \citet{Stickland87} to determine a radial velocity curve and an orbital solution for the primary component, but not to identify the secondary spectral lines. \citet{BGW92} used a tomographic technique on the same IUE data to separate the two components of the system. They found a mass ratio of $1.18 \pm 0.12$ (the secondary being more massive than the primary) and spectral types of O7.5~I and O6~I for the primary and the secondary, respectively. Given that they estimate an inclination around 71\degr, they found masses equal to 42.5~M$_{\odot}$ for the primary star and 51.0~M$_{\odot}$ for the secondary star. The secondary lines appeared rotationally broadened ($v\,\sin\,i = 310 \pm 20$~km~s$^{-1}$) but are visible throughout the orbit. Finally they found that the secondary contribution to the cross-correlation functions is relatively constant with orbital phase. \citet{Stickland97} also used IUE data and cross-correlation functions to study specifically the S-S effect in Plaskett's star. The conclusion was that the secondary radial velocities ''make no sense dynamically'', and that they are probably due to the presence of gaseous streams or envelopes in the system. However, \citet{BGR99} used the same method and concluded that there is no evidence of the S-S effect in \object{HD~47\,129}.

\begin{table}[t]
\caption{Journal of the observations of \object{HD~47\,129}. The first column indicates the instrument used to obtain the data, the second lists the Heliocentric Julian Date (HJD), the third provides the orbital phase ($\Phi$) as computed from our orbital solution (see Table \ref{tab:orbit:47129}) and the fourth and fifth ones give the radial velocities used to compute the orbital solution, respectively for the primary and secondary stars. The dispersion on the radial velocities are on average less than 2.0\,km\,s$^{-1}$ for the primary, and of the order of 20\,km\,s$^{-1}$ for the secondary component. }
\label{tab:journal:plaskett}
\centering
\begin{tabular}{c c c r r}
\hline
\hline
Instrument&HJD &$\Phi$ &\multicolumn{2}{c}{RV (km s$^{-1}$)}\\    
&$-$ 2 450 000 & &Primary&Secondary\\
\hline
\multicolumn{5}{c}{}\\
FEROS&2335.565& 0.28&$230.8$&$-149.2$\\
FEROS&2336.532&  0.35&$197.7$&$-141.0$\\
FEROS&2336.551& 0.35&$192.9$&$-119.7$\\
FEROS&2337.527& 0.42&$120.6$&$-117.8$\\
FEROS&2337.534& 0.42&$118.3$&$-118.2$\\
FEROS&2338.528& 0.49&$60.8$&$-29.2$\\
FEROS&2339.535& 0.56&$-28.4$&-\\
AURELIE&3648.675&0.49&-&-\\
AURELIE&3650.662&0.63&-&-\\
AURELIE&3652.673&0.77&-&-\\
AURELIE&3654.663&0.91&-&-\\
FEROS&3738.692 &0.74&$-169.0$&$134.5$\\
FEROS&3772.629 &0.10&$153.9$&$-151.4$\\
FEROS&3772.640 &0.10&$158.5$&$-129.7$\\
AURELIE&3775.486&0.30&-&-\\
AURELIE&3776.486&0.37&-&-\\
AURELIE&3777.478&0.44&-&-\\
AURELIE&3778.462&0.51&-&-\\
AURELIE&3779.318&0.57&-&-\\
AURELIE&3780.288&0.63&-&-\\
FEROS&3796.515& 0.76&$-171.7$&$124.9$\\
FEROS&3797.521& 0.83&$-152.0$&$145.2$\\
FEROS&3798.510& 0.90&$-80.9$&$147.1$\\
FEROS&3799.511& 0.97&$-23.2$&$74.6$\\
FEROS&3800.507& 0.04&$74.5$&$-$\\
AURELIE&3843.329&0.01&-&-\\
AURELIE&4033.700&0.24&-&-\\
AURELIE&4034.692&0.30&-&-\\
AURELIE&4035.672&0.37&-&-\\
SOPHIE&4162.432& 0.18&$205.9$&$-181.6$\\
SOPHIE&4164.307& 0.31&$219.7$&$-190.5$\\
SOPHIE&4168.468& 0.60&$-78.6$&$155.3$\\
SOPHIE&4172.305& 0.86&$-123.5$&$187.6$\\
SOPHIE&4174.333& 0.00&$31.1$&-\\
SOPHIE&4176.327& 0.14&$208.5$&$ -135.2$\\
SOPHIE&4182.321& 0.56&$-45.3$&$100.2$\\
SOPHIE&4184.350& 0.70&$-164.5$&$160.0$\\
\hline
\end{tabular}
\end{table}

\begin{figure}[t]
  \resizebox{\hsize}{!}{\includegraphics{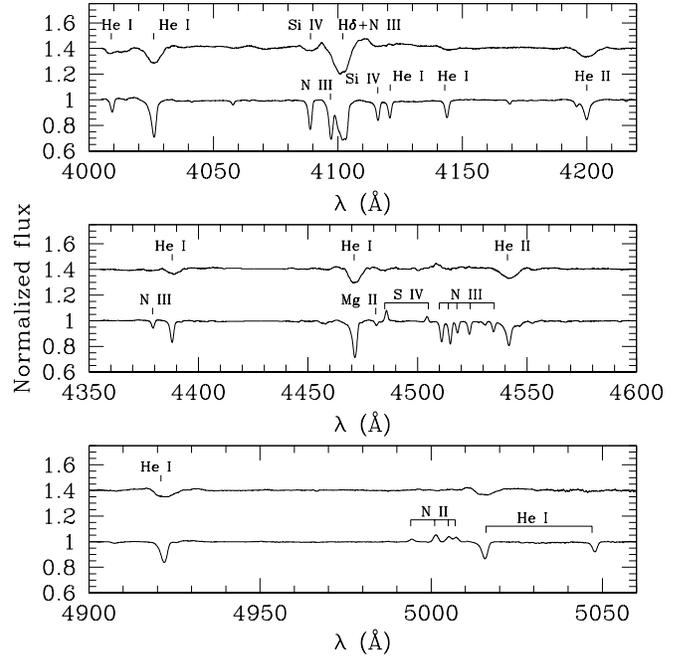}}
  \caption{Normalized disentangled spectra of \object{HD~47\,129}, over 3 distinct wavelength ranges. The secondary spectrum was vertically shifted by 0.4 units for clarity.}
  \label{fig:plaskett_disspec}
\end{figure}

Both components of Plaskett's star are very massive and luminous O-stars, and they thus both possess strong stellar winds that collide, producing X-ray emission in the interaction zone. An X-ray study of this object has been made by \citet{LRP06}, using new \textit{XMM-Newton} data and \textit{ROSAT} archival data. The RGS (Reflection Grating Spectrometers of the \textit{XMM-Newton} satellite) data showed a nitrogen overabondance by a factor of 7.44$^{+4.08}_{-2.59}$ in 2002 and 5.49$^{+4.51}_{-2.71}$ in 2003. We detected a modulation of the X-ray flux phased with the orbital motion, but could not find short term variability (from a few minutes to about one hour) that might have been expected from hydrodynamical instabilities of the wind-wind collision.

In this paper, we discuss a set of high resolution optical spectra of Plaskett's star, in order to constrain the evolutionary status of this peculiar system, and to determine the role played by the Struve-Sahade effect. Indeed, a better knowledge of the processes that take place in HD~47129 would be useful to understand the massive O+O binaries in general. The data have been obtained between 2002 and 2007 at the European Southern Observatory (ESO, Chile) and at the Observatoire de Haute-Provence (OHP, France). 

The paper is organized as follows. Section~\ref{observations} describes the observations and data reduction. In Section~\ref{rvcurve}, a radial velocity curve, computed with a disentangling method, and the corresponding orbital solution are presented. We discuss the spectral type of \object{HD~47\,129} in Section~\ref{spectraltype}, the behavior of the emission lines using a Doppler tomography technique in Section~\ref{emission_lines} and the behavior of the absorption lines in Section~\ref{sseffect}. 
We use a model atmosphere code in Section~\ref{martins} in order to fit theoretical spectra to the disentangled spectra of the components of Plaskett's star and discuss the evolutionary status of the system in Section~\ref{evolution}. Finally, Section~\ref{conclusions} summarizes the results and gives the conclusions.

\section{Observations and data reduction \label{observations}}

The journal of observations is presented in Table \ref{tab:journal:plaskett}. A part of the data was obtained by our team at ESO (La Silla) with the Fiber-fed Extended Range Optical Spectrograph (FEROS), mounted on the 1.5~m telescope (7 spectra in 2002, run~ESO~068.D-095(A)) and 2.2~m telescope (5 spectra in 2006, run~ESO~076.D-0294(A)). We have also added three FEROS archive spectra taken in 2006 (PI: Casassus, run~ESO~076.C-0431(A) and PI: Lo Curto, run~ESO~076.C-0164(A)). The wavelength domain of this echelle-spectrograph is [3550-9200]~\AA. Typical exposure times range from 5 to 15 minutes, with a mean signal to noise ratio (SNR) equal to 150 in 2002 and 340 in 2006. The spectral resolving power of FEROS is 48000. The data were reduced using an improved version of the FEROS pipeline \citep[see][]{SHR03} working in the MIDAS environment. 

The second part of the data was obtained in March 2007 at the OHP with the new echelle-spectrometer Sophie, mounted at the 1.93m telescope.  The Sophie instrument features a 52.65 grooves/mm R2 echelle grating and the spectra are projected onto an E2V~44-82~CCD detector. Thirty-nine spectral orders are currently extracted over a wavelength domain equal to [3872$-$6943]~\AA, with a resolving power of 40000 in the High-Efficiency (HE) mode \citep{BTST06}. The mean SNR in these spectra is equal to 290. The data reduction is done via an entirely automatic pipeline, adapted from the HARPS software designed at the Geneva Observatory.

Finally, data were also obtained with the Aur\'elie spectrometer, mounted at the 1.52~m telescope at OHP and equipped with a 2048~x~1024~CCD~EEV~42-20\#3, with a pixel size of 13.5~$\mu$m~squared. All spectra were taken with a 600~l/mm grating with a reciprocal dispersion of 16~\AA~mm$^{-1}$, allowing us to achieve a spectral resolving power of 8000 in the blue range (6 spectra in the [4450$-$4900]~\AA~domain), 9500 in the green range (7 spectra in the [5480$-$5930]~\AA~domain) and 11000 in the red range (1 spectrum in the [6340$-$6780]~\AA~domain). The spectra have a mean SNR equal to 300. The data reduction has been done with the MIDAS software as described in \citet{RDB04}. Because of their lower resolution, the Aur\'elie data were not used for the disentangling and Doppler tomography techniques applied in this paper (see Sections~\ref{rvcurve} and~\ref{emission_lines}). However, they provide a set of spectra well distributed in phase and we used them in order to check the accuracy of the results found with the FEROS and Sophie spectra.

\section{Orbital solution \label{rvcurve}}

As for the main sequence early-type binary systems studied in \citet{LRS07}, the radial velocities (RVs) were determined in an iterative way using a disentangling algorithm \citep{GL06} applied to three different wavelength domains of the FEROS and Sophie spectra ([4000$-$4220], [4360$-$4600] and [4900$-$5060]~\AA, presented in Fig.~\ref{fig:plaskett_disspec} in which the most important lines are labeled). We choose these domains because they present a lot of helium lines that are also useful for the spectral type determination, and exclude complicated emission lines such as \ion{He}{ii}~$\lambda$~4686 and H$\alpha$ (see Sect.~\ref{emission_lines}). For each of these domains, we built two specific masks\footnote{These masks are synthetic spectra equal to 0 except at the theoretical wavelengths of the lines where they are equal to 1.} for the cross-correlation with the primary and secondary spectrum, respectively. This mask is different for the primary and the secondary component because the nitrogen lines that are prominent in the primary spectrum are quasi absent from the secondary spectrum. The lines used for the RV calculation are H$\delta$; \ion{He}{i}~$\lambda\lambda$~4388,~4471,~4921, 5016; \ion{He}{ii}~$\lambda\lambda$~4200,~4542 and \ion{Si}{iv}~$\lambda$~4089. For the primary star, we also added \ion{He}{i}~$\lambda\lambda$~4009,~4026,~4120, 5048; \ion{Si}{iv}~$\lambda$~4116 and \ion{N}{iii}~$\lambda$~4097,~4196. The disentangling code converged to final separated spectra for each component and for each wavelength domain, and to corresponding radial velocities.

\begin{figure}[t]
\centering
\resizebox{0.9\hsize}{!}{\includegraphics{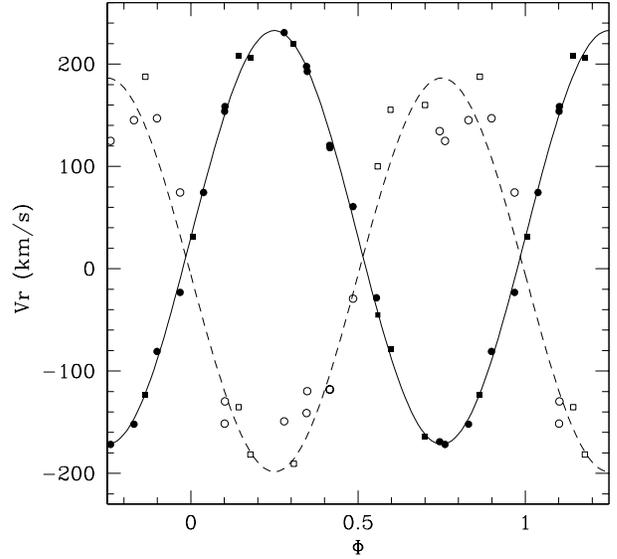}}
\caption{Radial velocity curve of \object{HD~47\,129}. Primary data are represented by filled symbols, while secondary data are represented by open symbols. Circles (resp. squares) stand for FEROS data (resp. Sophie data). The RVs have been determined through a disentangling of the primary and secondary spectra using the method of \citet{GL06}. $\Phi = 0$ corresponds to the phase of the conjunction with the primary being in front.}
\label{fig:vr_dis}
\end{figure}

\begin{table}[t]
\caption{Up: Orbital period of \object{HD~47\,129} calculated with the Fourier technique with all data available in the literature. Middle: Orbital solution of \object{HD~47\,129} computed from the RVs of the primary, obtained from the disentangling method. $T_0$ refers to the time of the primary conjunction ($\Phi = 0$). $\gamma$, $K$ and $a\,\sin\,i$ are respectively the systemic velocity, the semi-amplitude of the radial velocity curve, and the projected separation between the centre of the star and the centre of mass of the binary system. Quoted uncertainties are the $1\sigma$ error bars. Below: Derived elements of the secondary orbit (see text).}
\label{tab:orbit:47129}
\centering
\begin{tabular}{lrcl}
\hline	
\hline 
Parameter& & & \\
\hline
 & & & \\
$P$ (days)& 14.396257    &$\pm$ &0.000953   \\
\hline
 & & & \\
$e$ & 0     & &  (fixed) \\
$T_0$ (HJD)&2452331.547   &$\pm$&0.028  \\
$\gamma_\mathrm{P}$ (km s$^{-1}$)& 30.6     &$\pm$& 1.8 \\
$K_\mathrm{P}$ (km s$^{-1}$)&  202.2     &$\pm$& 2.6  \\
$a_\mathrm{P}$ $\sin\,i$ (R$_{\odot}$)&   57.5      &$\pm$& 0.7  \\ 
 F$_{\mathrm{mass,P}}$ (M$_{\odot}$)& 12.3     &$\pm$& 0.5  \\
rms (km s$^{-1}$)&       &8.6&\\
\hline
 & & & \\
 $\gamma_{\mathrm{S}}$ (km s$^{-1}$)& $-$6.0 &$\pm$& 9.6\\
 $K_\mathrm{S}$ (km s$^{-1}$)&  192.4     &$\pm$& 6.7  \\
 $M_{\mathrm{S}}\,\sin^3\,i$ (km s$^{-1}$)& 47.3 &$\pm$& 0.3\\
 $\frac{M_{\mathrm{S}}}{M_{\mathrm{P}}}$&  1.05 &$\pm$& 0.05\\
 
\hline
\hline
\end{tabular}
\end{table}

The primary RVs show very little scatter between the three domains and the dispersion on individual data points are on average less than 2.0\,km\,s$^{-1}$. For the secondary, the situation is unfortunately less clear and the dispersion is significantly larger (most of the time of the order of 20\,km\,s$^{-1}$). For three data points, the error exceeded 100~km~s$^{-1}$ and they have been completely excluded from the analysis. There are several reasons for this situation. One is the existence of changes in the line intensities of both the primary and secondary spectral lines. In particular, this leads to an imperfect subtraction of the spectral signature of the primary from the cross-correlation function, thus shifting the measured RV of the secondary toward the primary's RV. To circumvent this problem, we have scaled the primary spectrum by the ratio between the height of the main peak in the cross-correlation function evaluated at individual phases and the  equivalent in the cross-correlation function of the mean primary spectrum before subtracting the latter from the observed spectrum at the individual phases. This solves the issue for the primary line variability, but the line variations of the secondary still affect the resulting RVs (Fig.~\ref{fig:vr_dis}). Another reason for larger dispersion of RV's in the secondary star is the signifigant rotational broadening of the secondarys absorption lines, which makes it difficult to determine the centro\"{\i}d of the line. Table~\ref{tab:journal:plaskett} shows the averaved RV values from the three domains for both the primary and the secondary star. These values were then used one more time as a fixed input to our disentangling code to derive a final set of self-consistent disentangled spectra presented in Fig.~\ref{fig:plaskett_disspec}.

The second step consisted in a refining the orbital period by considering all the data available in the literature \citep{Struve48,Abhyankar1959,HC76,Stickland87,WG92,BB96} in addition to the spectra presented in this paper. The new orbital period, equal to $14.396257 \pm 0.000953$ days, has been determined with the generalized Fourier technique \citep{HMM85,GRR01}, especially adapted to unevenly spaced data. The error quoted here is empirical, and is equal to one tenth of the width of the peaks of the power spectrum. The theoretical formula of \citet{LS71} gives a smaller error value based on the signal to noise ratio but actually, with the inhomogeneous data set considered here, the empirical error is preferred.

The orbital solution was calculated with the program LOSP (Li\`ege Orbital Solution Package, see \citealt{RSG00}), keeping the orbital period fixed at the above value. This program is a modified version of the algorithm developed by \citet{WHS67}. The calculation was made on primary data only, as for an SB1 system and the solution, given in Table~\ref{tab:orbit:47129}, is close to the results found by \citet{Stickland87} with all the available data, and by \citet{Stickland97} with only the IUE data. A major difference between their solution and ours is the systemic velocity $\gamma$, which is $\sim 8$ km~s$^{-1}$ larger in our orbital solution. Our mass function is also higher than the mass function calculated with the IUE data.

From the SB1 solution, we have fitted the secondary radial velocities to the following relation: $V_{\mathrm{r,S}}(t) = c_1 V_{\mathrm{r,P}}(t) + c_2$ where $V_{\mathrm{r,P}}$ and $V_{\mathrm{r,S}}$ are respectively the primary and secondary radial velocity as a function of time, $c_1 = -\frac{M_{\mathrm{P}}}{M_{\mathrm{S}}}$ with $M_{\mathrm{P}}$, $M_{\mathrm{S}}$ the masses of the primary and secondary component respectively. We then found $Q = \frac{M_{\mathrm{S}}}{M_{\mathrm{P}}} = 1.05 \pm 0.05$, $K_{\mathrm{S}} = 192.4 \pm 6.7$~km~s$^{-1}$ and $\gamma_{\mathrm{S}} = -6.0 \pm 9.6$~km~s$^{-1}$. These results give $M_{\mathrm{P}}\,\sin^3\,i = 45.4 \pm 2.4$~M$_{\odot}$ and $M_{\mathrm{S}}\,\sin^3\,i = 47.3 \pm 0.3$~M$_{\odot}$. If we suppose an orbital inclination between 69.3 and 72.7\degr \citep{BGW92}, these numbers lead to masses between 52 and 55~M$_{\odot}$ for the primary component, and between 54 and 58~M$_{\odot}$ for the secondary component. These results are summarized in the second part of Table~\ref{tab:orbit:47129}.

Orbital solutions of individual spectral lines, based on direct measures performed on the observed spectra, have also been investigated and no significant differences have been noticed. These measures were only possible on the primary component of the system, the secondary lines being too faint to obtain reliable results with this method.

The projected rotational velocities ($v\,\sin\,i$) were derived using the Fourier technique \citep{SDH07,Gray2005}. For the primary, we found that the width of the \ion{He}{i}~$\lambda\lambda$~4471, 4921 and 5016 lines indicates $v\,\sin{i} =$~75, 62 and 60~km~s$^{-1}$, respectively. The \ion{He}{ii}~$\lambda\lambda$~4200, 4542 lines cannot be used here because of the severe blending with the nearby \ion{N}{iii} lines. As far as the secondary star is concerned, the \ion{He}{i}~$\lambda$~4471, 4921 and \ion{He}{ii}~$\lambda\lambda$~4200, 4542 lines yield $v\,\sin{i} =$~245, 230, 310 and 305~km~s$^{-1}$ respectively. Here, the latter values (\ion{He}{ii}) are likely to be the correct ones because the spectrum of the secondary displays some emission in the wings of the \ion{He}{i} lines which leads to apparently narrower \ion{He}{i} absorptions (see Table~\ref{tab:vsini}).

\begin{table}[t]
\caption{Projected rotational velocities, derived using the Fourier technique for some He~I and He~II lines. The errors have been estimated from the uncertainties on the position of the first zero in the Fourier transform of the observed profile.}
\label{tab:vsini}
\centering
\begin{tabular}{c| c c}
\hline	
\hline 
&\multicolumn{2}{c}{$v\,\sin\,i$ (km~s$^{-1}$)}\\
                          &Primary&Secondary \\
                          \hline
                          & & \\
He I $\lambda$ 4471&$75 \pm 5$&$245\pm 15$\\
He I $\lambda$ 4921&$62\pm 3$&$230\pm 15$\\
He I $\lambda$ 5016&$60\pm 3$&-\\
He II $\lambda$ 4200&-&$310\pm 20$\\
He II $\lambda$ 4542&-&$305\pm 20$\\
\hline
\hline
\end{tabular}
\end{table}

\section{Spectral type determination \label{spectraltype}}

A determination of the \object{HD~47\,129} spectral type can be made following Conti's criterion \citep[][and subsequent articles]{CA71}, which is based on the ratio between the equivalent width (EW) of \ion{He}{i}~$\lambda$~4471 and \ion{He}{ii}~$\lambda$~4542. We first measured the EWs directly on the observed spectra, after a standard deblending of both component lines by fitting two Gaussian profiles to the data when the stars are near quadrature phases. This leads to an O9 spectral type for the primary star and O6.5 for the secondary star. The uncertainties on these numbers correspond to a range for the primary between O8.5 and O9. On the other hand, the possible interval of spectral types for the secondary star is larger, between O6 and O7.5. This is mainly due to the presence of nitrogen lines which complicate the secondary component measure of \ion{He}{ii}~$\lambda$~4542. However, it is relatively close to the secondary's spectral type determined by \citet{BGW92} with their tomographic separation method, which is not the case for the primary component (they found O7.5~I~+~O6~I). The behavior of the \ion{He}{ii}~$\lambda$~4686 line suggests that at least one component of the system is not a main sequence star (Fig.~\ref{fig:4686}). The narrow emission line in the \ion{He}{ii}~$\lambda$~4686 complex seems indeed to move along with the primary star and would thus indicate a luminous giant or supergiant. We applied Conti's~O7$-$O9.5 luminosity criterion (log~W$^{\prime\prime}$~=~log~EW(\ion{Si}{iv}~$\lambda$~4089)~$-$~log~EW(\ion{He}{i}~$\lambda$~4143)) to the primary measures of EWs and it indeed yields a giant (III) luminosity class. 

Furthermore, we applied the same criteria on the disentangled spectra, and noticed substantial differences. Measures gave an O8~III spectral type for the primary spectrum and O7.5~I for the secondary spectrum. This is not surprising since Fig.~\ref{fig:plaskett_disspec} shows two very similar spectra, even though the secondary lines are heavily broadened. 

Comparison with the criteria from the atlas of \citet{WF90} (\ion{He}{i}~$\lambda$~4771/\ion{He}{ii}~$\lambda$~4542 and \ion{He}{i}~$\lambda$~4026/\ion{He}{ii}~$\lambda$~4200 EW ratios) further confirms the O8 spectral type of the primary star with a huge enhancement of the nitrogen lines. The ratio \ion{He}{i}~$\lambda$~4338/\ion{He}{ii}~$\lambda$~4542 in the primary spectrum argues against a luminosity class brighter than III, however it is interesting to note that the emissions seen in \ion{S}{iv}~$\lambda\lambda$~4486,~4504 seem to be more typical of a supergiant\footnote{We note however, that the ''selective'' nature of these lines is not fully established \citep{Walborn2001,WR01}.}. Most criteria thus indicate O8~III/I for the primary star. Concerning the secondary star, the \ion{He}{i}~$\lambda$~4771/\ion{He}{ii}~$\lambda$~4542 ratio in the disentangled spectra clearly shows a spectral type later than O7, while the absence of \ion{S}{iv} emission lines suggests that it has a luminosity class III. Finally, the secondary should be of O7.5 spectral type, rather than O8, because of the absence of \ion{He}{i}~$\lambda\lambda$~4143,~4388. In conclusion, we favor the following spectral types: O8~III/I for the primary, O7.5~III for the secondary.

We derived an optical brightness ratio, by comparing the EW ratio between the primary and secondary disentangled spectra and the same ratio evaluated for a sample of presumably single stars of the same spectral type \citep{CA71,Conti73}. The comparison was performed for the \ion{He}{i}~$\lambda\lambda$~4026,~4388,~4471 and \ion{He}{ii}~$\lambda\lambda$~4200,~4542 lines, and we found $\frac{L_{\mathrm{P}}}{L_{\mathrm{S}}}=1.9 \pm 0.1$. The primary star is thus brighter than the secondary star in the optical domain. This result is similar to the optical brightness ratio found by \citet{BB96} with their tomography algorithm. 

From the average of the $UBV$ photometric data compiled by \citet{Reed2005}, we obtain $V = 6.06 \pm 0.01$ and $B - V = 0.04 \pm 0.03$. Assuming $(B - V)_0 = -0.27$ \citep{MP06}, we infer $E(B - V) = 0.31 \pm 0.03$, which implies $A_{\mathrm{V}} = 0.96 \pm 0.09$ if we adopt $R_{\mathrm{V}} = 3.1$. Assuming \object{HD~47\,129} to be a member of the Mon~OB2 association, the distance modulus should be about 10.9 \citep{Humphreys1978}. Hence, the absolute magnitude of \object{HD~47\,129} would be $M_{\mathrm{V}} = -5.80 \pm 0.09$. From the optical brightness ratio inferred above, we then arrive at absolute magnitudes of $M_{\mathrm{V,P}} = -5.342 \pm 0.105$ and $M_{\mathrm{V,S}} = -4.645 \pm 0.131$ for the primary and secondary respectively. The value for the primary is somewhat (0.13 magnitude) fainter than typical O8~III $V$-band luminosities \citep{MP06} whilst the $M_{\mathrm{V}}$ of the secondary is about 0.9 magnitude fainter than the typical O7.5~III magnitude and would be much more typical of a main sequence or subgiant luminosity class. A better agreement between the absolute magnitude and the spectral type can be obtained, at least for the primary star, by increasing the distance of the system. Indeed, if we suppose that the primary star has an absolute magnitude equal to a typical value of an O8~III~star ($M_{\mathrm{V,P}} = -5.47$, \citealt{MP06}), the distance modulus would be equal to 11.0, which is still compatible with the distance determined for other stars belonging to the Mon~OB~2 association (i.e.~\object{HD~46\,202} and \object{HD~47\,240}, \citealt{Humphreys1978}). Taking this new distance into account, the absolute magnitude of the secondary star is now equal to $-4.72$, which is still too faint for an O7.5~III star and is more typical of an O6.5$-$O7~V star. If we adopt bolometric corrections of $-3.16$ and $-3.25$ for the primary and secondary respectively \citep{MP06}, we infer bolometric luminosities of  $\log{L/L_{\odot}} = 5.35 \pm  0.04$ and $5.09 \pm 0.04$  respectively. 

At this stage, it is worth emphasizing that the rapid rotation of the secondary (see~Section~\ref{rvcurve}) could lead to a rotational flattening that would make this star appear hotter near the poles than at the equator. As a result, the spectral type determined from the \ion{He}{i}/\ion{He}{ii} line intensity ratios might actually be biased. We applied the equations given by \citet[][Appendix A]{Maeder1999}, assuming that the inclination of the rotation axis is the same as the orbital inclination (i.e. 71\degr), and that the angular velocity is constant. We then found that, for an equator radius equal to 14.1 R$_{\odot}$ (which is a typical value of an O7.5~III star, \citealt{MSH05}), the polar radius is equal to 13.2 R$_{\odot}$. The von Zeipel theorem applied to our case indicates a difference in effective temperature equal to $\sim$~2\,500~K between the pole and the equator \citep{vonZeipel1924}. This means that the temperature of the secondary's pole would be equal to 35\,500~K (since we derive $T_{\mathrm{eff}}$~=~33\,000~K, see Sect.~\ref{martins}), which is compatible with an O7~V$-$O7.5~V star. The secondary star of \object{HD~47\,129} could thus be an O7~V star that shows an apparent O7.5~III spectrum because of its rapid rotation. 

In conclusion, the discrepancy between the spectral type of the primary star and its absolute luminosity can be solved by adopting a distance modulus of 11.0 for the system. For the secondary star this solution is not sufficient, but the rest of the discrepancy could be due to the large rotational velocity of the star, that would induce a strong gravity darkening leading to a lower temperature (and hence apparently later spectral type) over the equatorial regions. The same effect also yields a lower gravity near the equator than close to the poles, hence leading to an apparent giant luminosity class. 

Nevertheless, the rather modest luminosities are still surprising given the minimum dynamical masses of the stars inferred from our orbital solution (45 and 47~M$_{\odot}$, see~Sect.~\ref{rvcurve}): both stars would appear underluminous for their dynamical masses. The best estimate of the inclination is 70\degr \citep{BGW92}, leading to masses of 55 and 57~M$_{\odot}$, and thus enhancing the discrepancy. Actually, this problem can be formulated in a different way: the stars have spectral types that are too late for their dynamical masses, which are more typical of giant or main sequence O3 stars. Increasing the distance would imply that Plaskett's star does not belong to the Mon OB2 association, which is unlikely. Another still unknown effect might thus be present here.

\section{The emission lines and Doppler tomography \label{emission_lines}}

\begin{figure}[t]
\centering
\resizebox{\hsize}{!}{\includegraphics{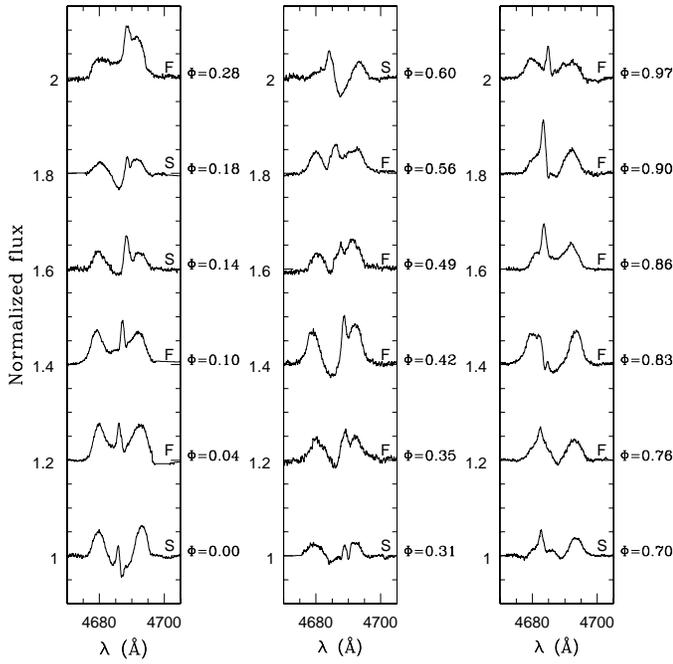}}
\caption{\ion{He}{ii}~$\lambda$~4686 line in the spectra of \object{HD~47\,129} from FEROS (F) and Sophie (S) observations. The orbital phases are indicated to the right of each panel.}
\label{fig:4686}
\end{figure}

\begin{figure}[t]
\centering
\resizebox{\hsize}{!}{\includegraphics{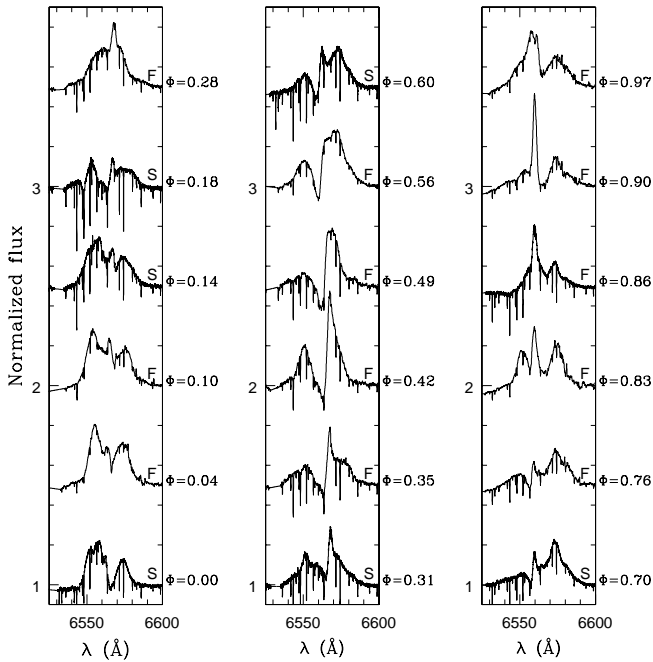}}
\caption{Same as Fig~\ref{fig:4686}, but for the H$\alpha$ line in the spectra of \object{HD~47\,129}.}
\label{fig:halpha}
\end{figure}

\object{HD~47\,129} displays prominent phase-locked profile variability of the strongest  emission lines (H$\alpha$, \ion{He}{ii}~$\lambda$~4686, \ion{He}{i}~$\lambda$~5876) in its optical spectrum. As becomes clear from a montage of the observed profiles (Figs.~\ref{fig:4686} and~\ref{fig:halpha}), the H$\alpha$ and \ion{He}{ii}~$\lambda$~4686 lines consist of several components: a narrow emission line that roughly follows the orbital motion of the primary star, and two broad emissions (a blue-shifted and a red-shifted one) that do not seem to move with the orbital cycle. 

Further insight into the origin of these variations can be obtained from the Doppler tomography technique which allows to map the formation region of an emission line in velocity space. Here we use an implementation of the Doppler tomography based on a Fourier filtered back projection algorithm \citep{Horne1991} already used by \citet{RCE02,RCdB05} for the analysis of line profile variations of \object{HDE~228\,766} and \object{WR~20a}. For this purpose, we adopt a reference frame centered on the centre of mass of the binary with the $x$-axis pointing from the primary to the secondary and the positive $y$-axis pointing along the direction of the secondary's orbital motion. The Doppler tomography technique assumes that the phase dependence of the radial velocity $v(\Phi)$ of any gas flow that is stationary in the rotating frame of reference of the binary can be described by a so-called `S-wave' relation:

\begin{equation}
v(\Phi) = v_x\,\cos{(2\,\pi\,\Phi)} - v_y\,\sin{(2\,\pi\,\Phi)} + v_z
\label{eq1}
\end{equation}
where $\Phi$ is the orbital phase ($\Phi = 0$ at conjunction with the primary being in front), whilst $v_x$ and $v_y$ reflect the projected velocity components along the $x$ and $y$ axes and $v_z$ corresponds to the apparent systemic velocity of the line under investigation. A Doppler map consists of a projection of the S-wave relation of Eq.~\ref{eq1} onto the $(v_x, v_y)$ plane (for which $v_z = \gamma$). Each pixel in a Doppler map, specified by its velocity coordinates is associated with a particular S-wave (see e.g. \citealt{Horne1991} for a detailed discussion of the method).

\begin{figure}
\begin{center}
\resizebox{8cm}{!}{\includegraphics{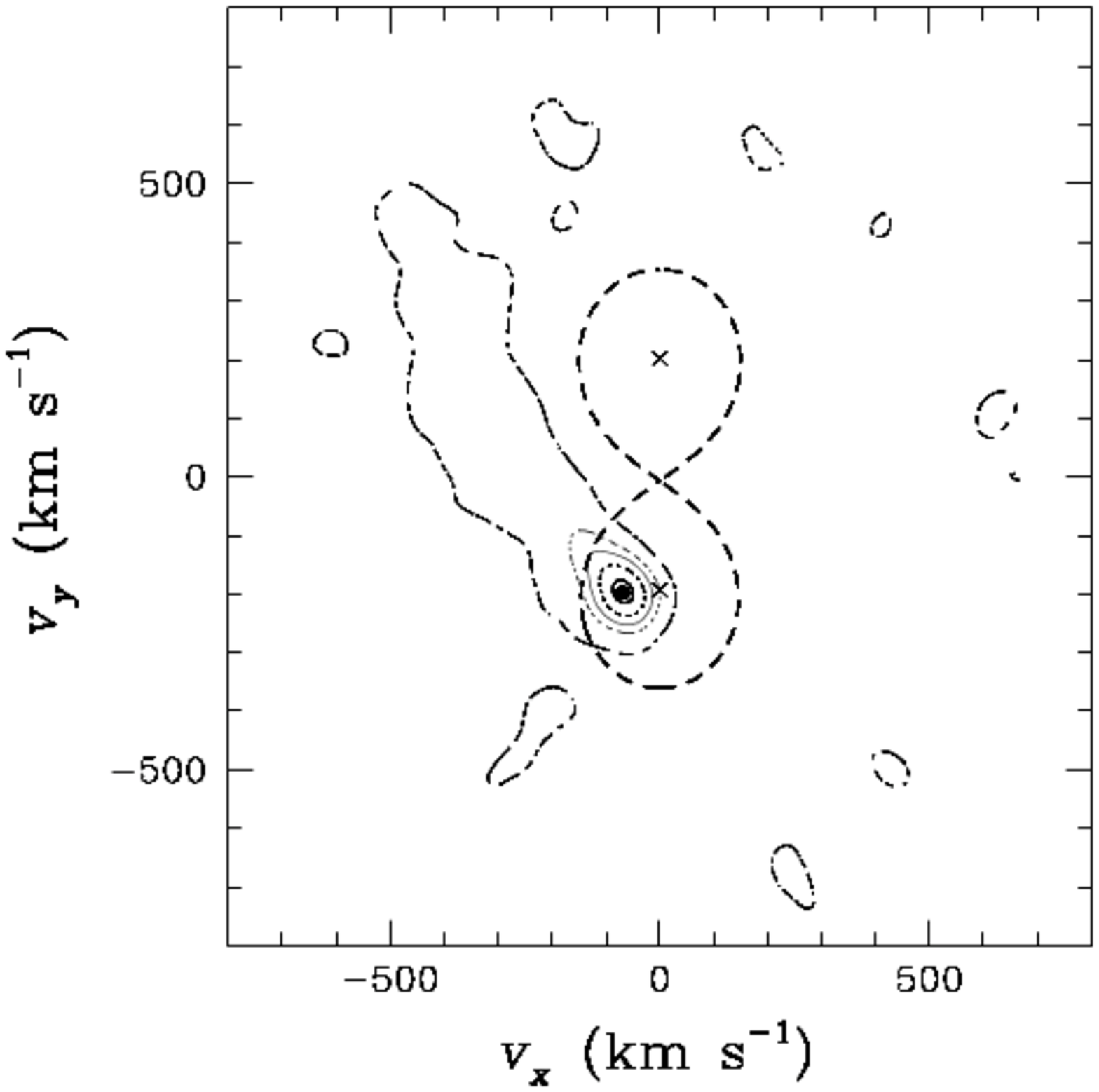}}
\resizebox{8cm}{!}{\includegraphics{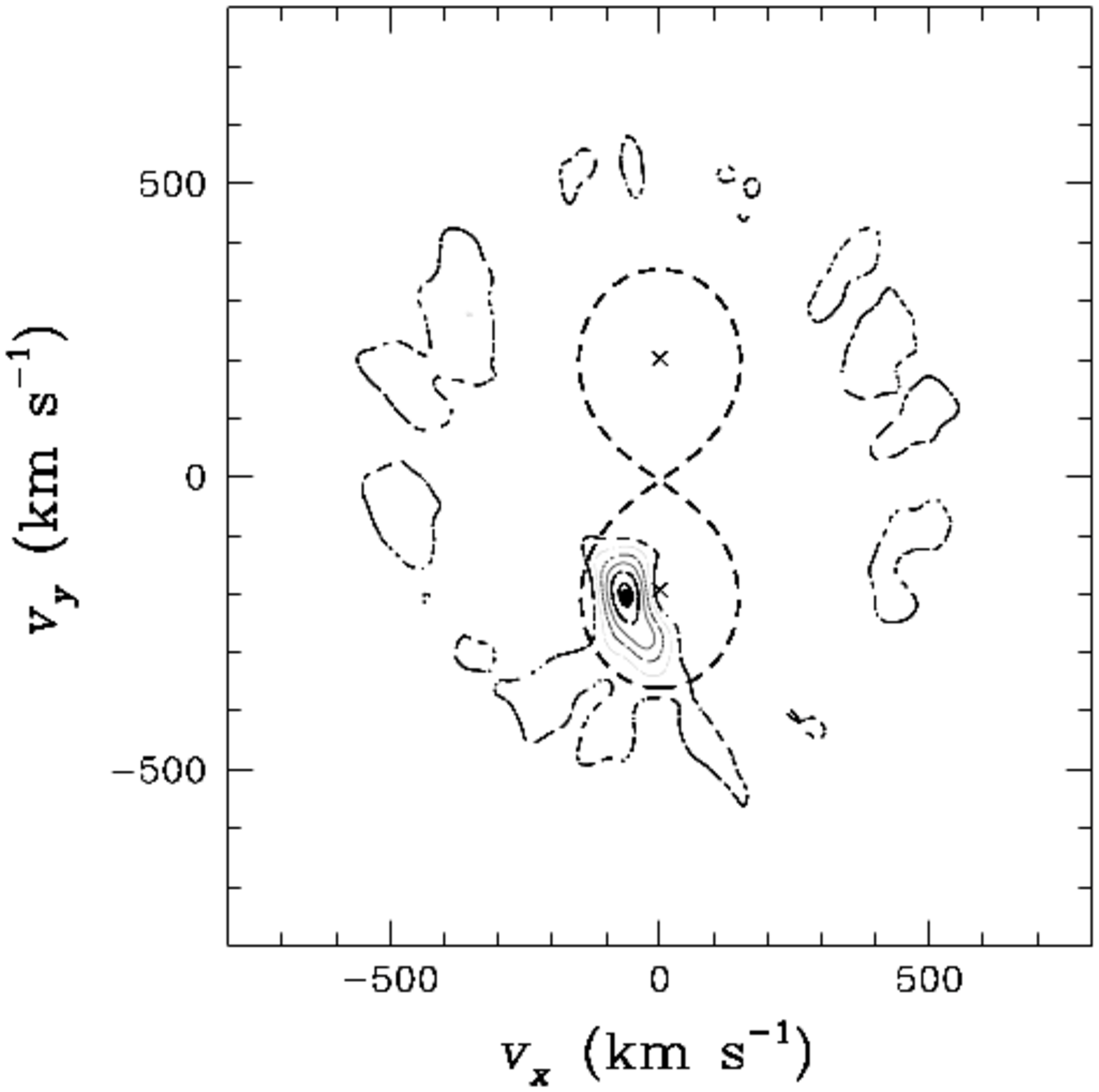}}
\caption{\label{doppler} Top: Doppler map of the H$\alpha$ emission line in the spectrum of \object{HD~47\,129}. The crosses correspond to the radial velocity amplitudes of the centre of mass of the primary and secondary. The shape of the Roche lobe in velocity-space (thick dashed line) has been calculated for a mass ratio (secondary/primary) of 1.05. The Doppler map was computed with $v_z$ set to $+30$~km~s$^{-1}$. The black dot indicates the position of the highest peak and the contours correspond to levels of 0.95, 0.80, 0.65, 0.50, 0.35 and 0.20 times the maximum emissivity. Bottom: Same, but for the \ion{He}{ii}~$\lambda$~4686 emission.} 
\end{center}
\end{figure}

To build the Doppler maps of the \ion{He}{ii}~$\lambda$~4686 and H$\alpha$ emission lines (Fig.~\ref{doppler}), the data were weighted so as to account for the uneven sampling of the orbital cycle. The Doppler maps of both lines present their highest peak near the centre of mass of the primary with a slightly negative $v_x$ ($-$60.0 and $-$68.0~km~s$^{-1}$ for \ion{He}{ii}~$\lambda$~4686 and H$\alpha$ respectively) and $v_y$ very close to the amplitude $K_{\mathrm{P}}$ of the primary's radial velocity curve (202.5~km~s$^{-1}$). This peak is associated with the narrow emission component visible in both lines and discussed above. In addition, both Doppler maps reveal a roughly annular region with patchy emission structures. This annulus is most prominent for the \ion{He}{ii}~$\lambda$~4686 line. The structure is centred on a point with coordinates $(v_x, v_y) = (0, +25)$~km~s$^{-1}$, with inner and outer radii of about 400 and 600~km~s$^{-1}$ respectively. This feature is responsible for the two broad, roughly stationary emissions discussed above. In the H$\alpha$ Doppler map, the structure appears more elongated and less well defined. In addition, the H$\alpha$ Doppler map shows an extended `emission bridge' connecting the peak at $(v_x, v_y) = (-68, -198)$~km~s$^{-1}$ with the annulus near $(v_x, v_y) \sim (-450, +450)$~km~s$^{-1}$. 

Let us now turn to the possible interpretations of the various features seen in the velocity maps. First of all, we consider the peaks of the maps that occur at  slightly negative $v_x$ with respect to the motion of the primary star. The velocity coordinates of this emission region hence indicate a motion from the secondary toward the primary. A possible interpretation could thus be that this emission arises from a region where the primary wind is compressed by the interaction with the secondary wind (which is flowing toward the primary). In terms of a colliding wind interaction, this would correspond to the head of the shock which should indeed be displaced toward the primary. The narrow emission line component thus likely forms near the region where most of the X-rays are emitted \citep{LRP06}. This scenario is in agreement with the one proposed by \citet{WG92}. Measures of radial velocities performed on the FEROS, Sophie and Aur\'elie spectra also showed that the narrow emission lines come from a region of the wind close to the primary star. Actually, both S-waves are similar to the primary's RV curve presented in Fig.~\ref{fig:vr_dis}, but with a lag in phase of 0.04 for \ion{He}{ii}~$\lambda$~4686 and 0.06 for H$\alpha$.  This result corresponds to the colliding wind model of \citet{WG92} but with an emission zone of H$\alpha$ maybe a little closer to the system's centre of mass than the emission zone of \ion{He}{ii}~$\lambda$~4686. 

An alternative explanation for the narrow component could be a 'hot spot' due to the impact of a gas stream from a Roche lobe filling component on its companion. However, \object{HD~47\,129} is currently likely in a post Roche lobe overflow (RLOF) configuration where the primary transferred material to the secondary \citep{BGW92}. The $(v_x, v_y)$ components of the narrow emissions are therefore at odds with such a scenario.  

\begin{figure}[htb]
\begin{center}
\resizebox{6cm}{!}{\includegraphics{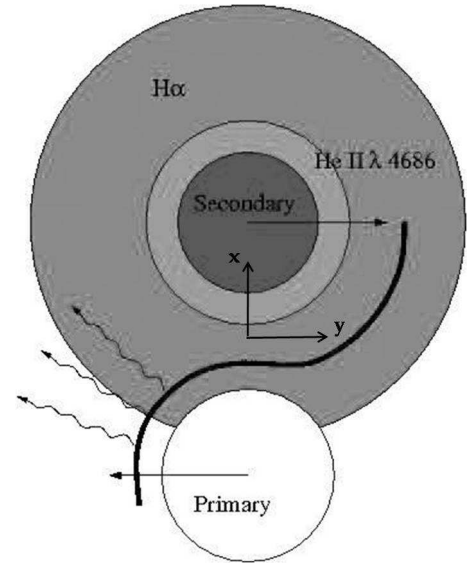}}
\caption{\label{sketch} Schematic pole-on view of the interaction in \object{HD~47\,129} (not to scale). The sense of the motion of the two stars is given by the arrows. The secondary star has a rotationally flattened wind. The \ion{He}{ii}~$\lambda$~4686 and H$\alpha$ emission regions are highlighted and the wake due to the motion of the primary across the secondary wind is indicated by the thick curve. The X-ray emitting region near the surface of the primary is also indicated (undulated arrows).} 
\end{center}
\end{figure}

A striking feature is the annular-like structure seen in the Doppler map that results from the roughly stationary double-peaked feature in the spectrum. Actually, both peaks display some variation in intensity and radial velocity, but they cannot be directly linked to the orbital phase of the system (this phenomenon is also observed in the absorption lines, see Sect.~\ref{sseffect}). This kind of structure could reflect the existence of an accretion disk around one of the stars. Indeed, it is believed that Plaskett's star has undergone mass exchange through a RLOF of the primary component in the past. However, it seems unlikely that this interaction would still be ongoing, especially since several features rather hint at material flowing from the secondary toward the primary (see above), i.e.~in the opposite sense compared to what is expected for the RLOF of the primary. Another point concerns the fact that the annular structure does not appear to be centered on any of the two stars. A more likely explanation could be that the secondary star has a wind flattened by its extremely high rotational velocity and hence compressed into its equatorial plane in a way similar to the O9.5~V star \object{HD~93\,521} \citep{BIT94}. The outer regions of such a flattened wind, where the H$\alpha$ line forms, would probably be heavily disturbed by the presence of the primary star orbiting around the secondary. However, the inner regions of the wind, where the \ion{He}{ii}~$\lambda$~4686 line forms, would be less affected, thus explaining why the feature is best seen in the Doppler map of the latter line. We performed numerical computations that showed that it should be possible to qualitatively reproduce this annulus with a combination of the wind's rotation velocity and the wind's expansion velocity.  However, it was not possible to obtain exactly the same annulus because the variation of the wind's rotation velocity with the distance from the secondary star as well as the influence of the UV primary radiation, which is supposed to reduce the secondary wind velocity, were unknown. 

As we have seen in Section~\ref{spectraltype}, the primary star is intrinsically more luminous than the secondary star and should hence have the stronger wind. Furthermore, the analysis of ROSAT-HRI archive data and of XMM Newton data have shown that a decrease in the X-ray flux seems to occur when the primary star is in front, which also indicates a stronger primary's wind, absorbing more the X-ray flux than the secondary's wind \citep{LRP06}. On the other hand, both the estimation of the component's mass loss rate in \citet{LRP06} and the optical emission lines suggest that the secondary star has the stronger wind and thus that the wind interaction zone is displaced toward the primary star. This apparent contradiction can be solved easily if the bulk of the secondary's wind is indeed concentrated in the equatorial plane as suggested above.
\begin{figure*}
\centering
\subfigure[\ion{He}{i}~$\lambda$~4026]{\resizebox{0.24\hsize}{!}{\label{fig:4026:028}\includegraphics{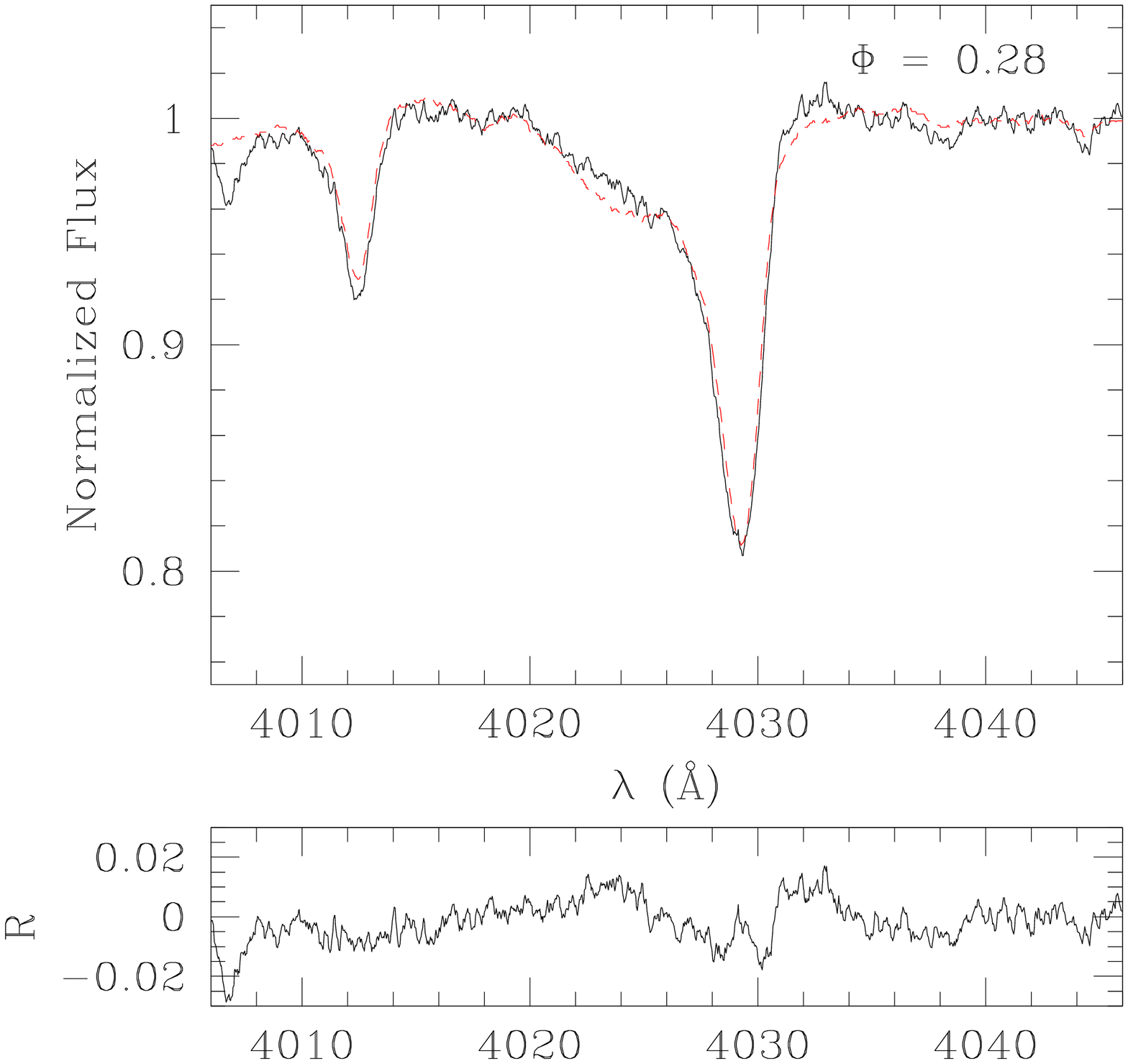}}}
\subfigure[\ion{He}{i}~$\lambda$~4026]{\resizebox{0.24\hsize}{!}{\label{fig:4026:074}\includegraphics{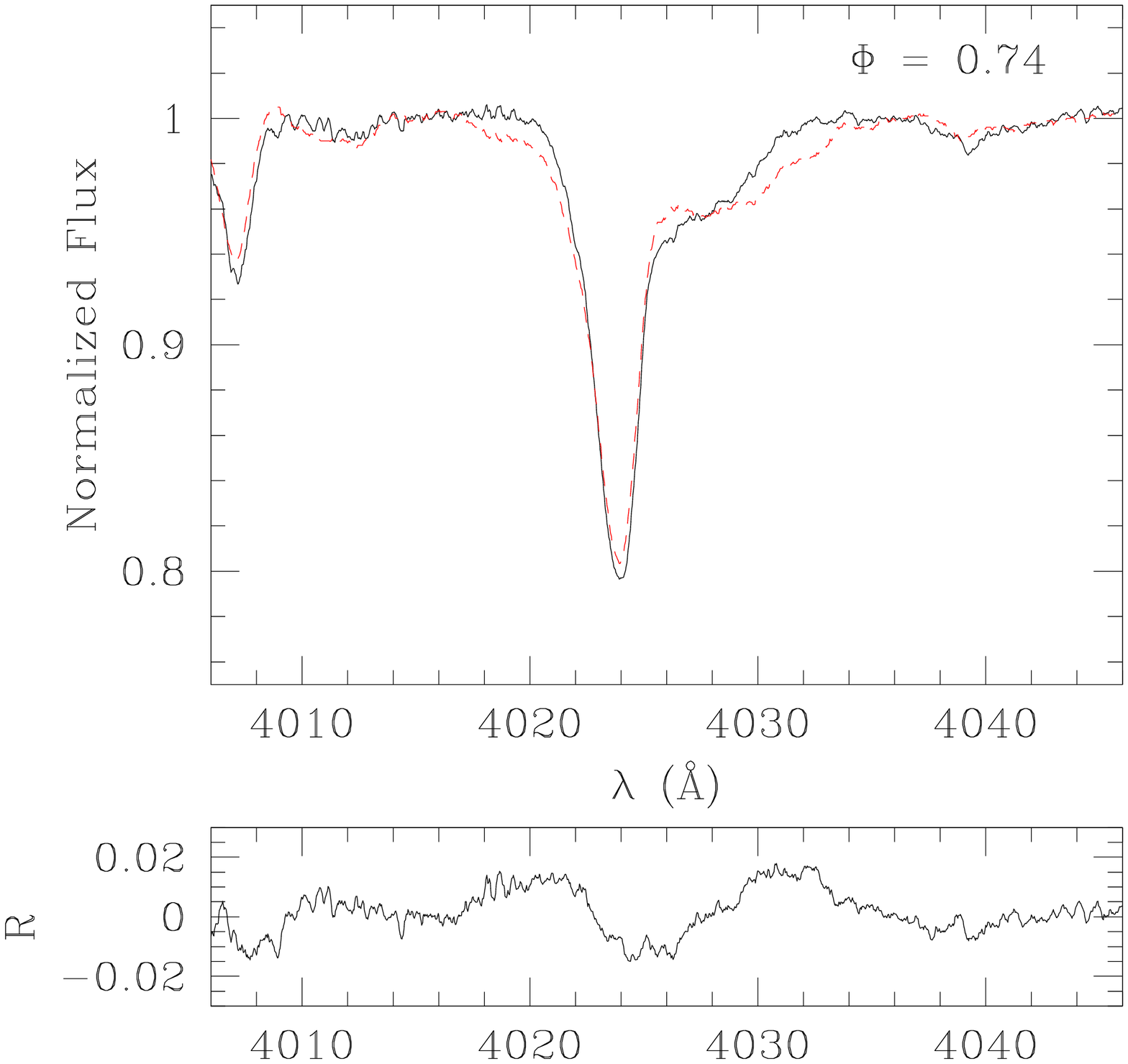}}}
\subfigure[\ion{N}{iii}~$\lambda$~4197 and \ion{He}{ii}~$\lambda$~4200]{\resizebox{0.24\hsize}{!}{\label{fig:4200:028}\includegraphics{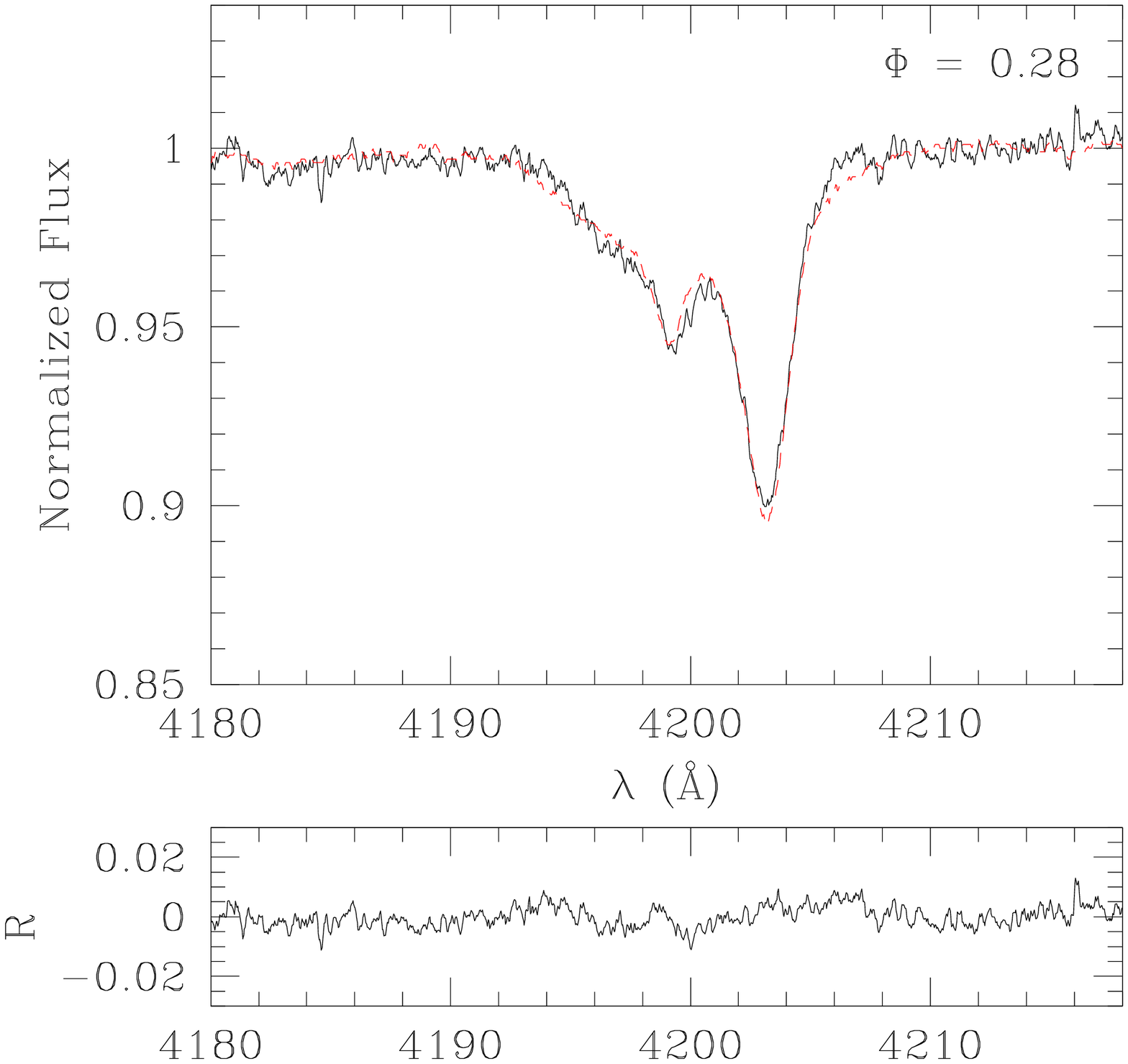}}}
\subfigure[\ion{N}{iii}~$\lambda$~4197 and \ion{He}{ii}~$\lambda$~4200]{\resizebox{0.24\hsize}{!}{\label{fig:4200:070}\includegraphics{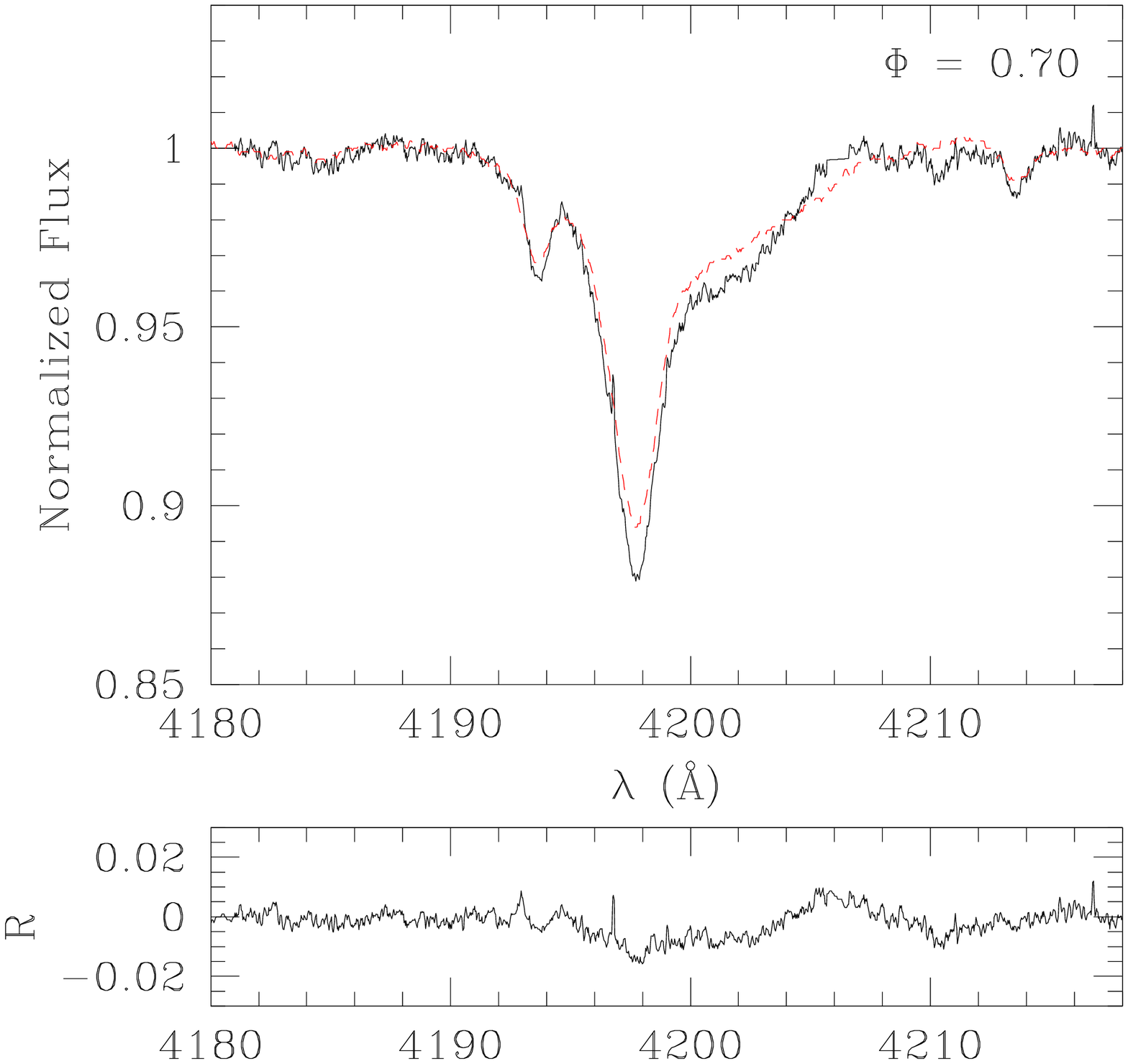}}}
\caption{Observed FEROS (a, b and c) and Sophie (d) helium lines of Plaskett's star (solid line) and disentangled spectra appropriately shifted in phase (red dotted lines). The lower panels yield the residuals (observed-reconstructed) in units of the normalized spectrum.}
\label{fig:sseffect}
\end{figure*}

Finally, the bridge-like structure seen in the H$\alpha$ map is most likely produced by the wake created by the primary star moving across the flattened wind of the secondary star. 

A schematic view of a possible wind interaction between the components of Plaskett's Star, where the secondary wind is flattened by the action of its rapid rotation, is shown in Fig.~\ref{sketch}.

\section{The absorption lines and the Struve-Sahade effect\label{sseffect}}
The S-S effect is known to appear in massive early type O~+~O binaries and impacts their fundamental parameters, such as masses and luminosity ratios. A systematic study of this particular effect has been made by our team on main sequence systems \citep{LRS07}, and we have determined that the S-S effect did not affect all studied lines of the spectra and that in most cases both stars of the system were concerned by the phenomenon. In order to continue the study, we analyse here the optical spectra of Plaskett's star which is one of the giant/supergiant O-type binaries that have been reported to display the Struve-Sahade effect.

Measures of equivalent widths made on all observed spectra (with FEROS, Aur\'elie and Sophie) showed first that, even with high resolution and reasonable S/N ratio data, separating the primary and secondary contribution to a given line is not a simple task. Actually, the secondary spectrum is so broadened and sometimes so faint that the measures show a large dispersion. The method applied in \citet{LRS07} was thus not optimal to check for the presence of the S-S effect in \object{HD~47\,129}. Indeed, the standard deblending of both components by fitting two gaussians on the data showed that, among all tested lines (\ion{He}{i}~$\lambda\lambda$~4026,~4388,~4471,~4713,~4921,~5016; \ion{He}{ii}~$\lambda\lambda$~4200,~4542 and \ion{N}{iii}~$\lambda\lambda$~3999,~4003,~4195,~4379), only two absorption lines seemed to be slightly stronger at $\Phi = 0.75$ (when the secondary star is receding): \ion{He}{i}~$\lambda$~4026 and \ion{He}{i}~$\lambda$~4388.

On the other hand, the disentangled spectra can be considered as a mean of the observed spectra. We can thus check for the presence of the S-S effect by comparing these mean profiles to the observed ones for each specific orbital phase. For this purpose, we have reconstructed the spectra by shifting the disentangled profiles of the primary and secondary by the appropriate radial velocities of these components. In this way, we find that at specific orbital phases, the observed line profiles often differ significantly from the reconstructed  ones. The strongest variations are clearly associated with the secondary star's lines that vary both in width and intensity. However, to a lesser extent, some variations are also observed for the primary star's lines (see e.g. the \ion{N}{iii}~$\lambda$~4197 line in Fig.~\ref{fig:sseffect}). It is not completely clear to what degree the variations of the absorption lines are phase-locked. Indeed, whilst some observations taken at close phases exhibit similar deviations from the reconstructed profiles, other observations with almost identical phases reveal totally different profiles. As a result, the line profile variations appear dominated by a modulation that is either stochastic or at least not tied to the orbital phase.

As stated above, the variations mostly affect the lines of the secondary star and the strongest deviations are seen in the wings of the \ion{He}{i} lines. On some spectra, the observed profile appears broader than the average, whilst on other data the reverse situation is observed. There are a number of possible scenarios to explain variations that are not linked to the orbital cycle. The fact that the variability mostly affects the wings of the secondary's \ion{He}{i} lines could be a hint that it actually stems from variations of weak emission wings. Indeed, such wings are clearly seen to flank some of the stronger \ion{He}{i} lines (see Fig.~\ref{fig:plaskett_disspec}). In the context of our suggestion that the secondary of \object{HD~47\,129} is surrounded by a flattened wind, the variability of these emission wings could then reveal a rather complex velocity structure of this flattened outflow. If correct, this interpretation would imply that the apparent variations of the absorption lines in \object{HD~47\,129} actually result from the effect of the turbulence and other velocity perturbations on the emissions formed in the flattened wind of the secondary star. Alternatively, the variability could be tied to the rotational cycle of the secondary or to non-radial pulsations of the latter. Concerning the first possibility, we note that the uncertainties on the rotational period of the secondary prevent us from further exploring this possibility. Non-radial pulsations in rapidly rotating O~stars can induce prominent line profile variability (see e.g. the case of \object{HD~93\,521}, e.g. \citealt{HR93}). However, the amplitude of the line profile variations in  \object{HD~47\,129} are rather substantial (several percent of the combined continuum, compared to less than one percent for \object{HD~93\,521}) and accounting for the brightness ratio between the components of Plaskett's star would imply intrinsic amplitudes even about three times larger.

\section{Modeling the spectra with an atmosphere code \label{martins}}

We used the model atmosphere code CMFGEN \citep{HM98} in order to derive the wind and photospheric properties of both components of \object{HD~47\,129} from the disentangled spectra. CMFGEN calculates non-LTE atmosphere models that include winds and the treatment of line-blanketing. Its main characteristics and hypothesis can be found in \citet{MHP08}. In addition to our optical spectra, we made a comparison between IUE archive spectra to which we have applied our disentangling method and theoretical spectra in order to determine the wind parameters. We first performed the tests taking the luminosities we observed for the stars into account, and then  luminosities equal to the average value for their spectral type (O8~III/I~+~O7.5~III, \citealt{MSH05}). The results provided by these two models are not significantly different, and thus fall within the same uncertainties.

\begin{table}
\caption{Some parameters of Plaskett's star, determined from simulations with the atmosphere code CMFGEN. T$_{\mathrm{eff}}$ stands for the effective temperature, $\dot{M}$ for the mass loss rate, $f$ for the filling factor and $v_{\infty}$ for the wind terminal velocity. The abundances of the listed elements are in number (relative to H), and compared to the solar abundances of \citet{GAS07}.}
\label{table:cmfgen}
	\centering
		\begin{tabular}{l c c}
		\hline
		\hline
		      &Primary&Secondary\\
		\hline
		    & & \\
		Spectral type & O8 III/I & O7.5 V/III \\
	$\log{L/L_{\odot}}$& 5.35 (fixed) & 5.09 (fixed)\\
		T$_{\mathrm{eff}}$ (K)& $33500 \pm 2000$& $33000\pm 2000$\\
		
		$\log\,g$&$3.5 \pm 0.1$& $3.5 \pm 0.1$\\
		
		He/H& $0.10 \pm 0.03$&$0.15 \pm 0.05$ \\
		
		He/He$_\odot$&$1.17 \pm 0.35$&$1.76 \pm 0.53$\\
		
		C/H&$5.0\times10^{-5} \pm 1.5\times10^{-5}$ & $2.45\times10^{-4}$ (fixed)\\
		C/C$_\odot$&$0.2\pm0.1$&$1.0$ (fixed)\\
		
		N/H&$1.0\times10^{-3}\pm 0.3\times10^{-3}$ &$1.0\times10^{-5}\pm 0.3\times10^{-5}$ \\
		N/N$_\odot$&$16.6 \pm 5.0$&$0.2 \pm 0.1$\\
		
		Mg/H&$1.9\times10^{-5}\pm 0.6\times10^{-5}$ & -\\
		Mg/Mg$_\odot$&$0.6 \pm 0.2$&-\\
		
		$\dot{M}$(M$_{\odot}$ yr$^{-1}$)&$1.7\times10^{-7} \pm 0.3\times10^{-7}$&-\\
		$f$&$0.05 \pm 0.03$&-\\
		$v_{\infty}$ (km s$^{-1}$)&2450& - \\
		\hline
		\hline
		\end{tabular}
\end{table}

Table~\ref{table:cmfgen} gives the main results we obtained. The effective temperatures and the effective gravity are compatible with the values given by \citet{MSH05} for the spectral types we found in Section~\ref{spectraltype}. The IUE spectra allowed us to calculate the wind parameters of the primary star. The $f$ parameter, or filling factor, is the volume occupied by matter in the wind divided by the total volume of the wind. Its value has been adopted in order to best reproduce the \ion{N}{iv}~$\lambda$~1720 and \ion{O}{v}~$\lambda$~1371 lines, which are particulary sensitive to clumping. The mass loss rate we found here is three times weaker than the theoretical mass loss rates given by \citet{LRP06}. The uncertainty on the quantity $\log\,\frac{\dot{M}}{\sqrt f}$ is about 0.2$-$0.3 dex and was estimated as the change beyond which the corresponding synthetic spectrum does not fit anymore the observed spectrum. This can be considered as a 3 sigma uncertainty. The abundances are listed by number relative to hydrogen and the corresponding error bars are equal to $\sim 30$\%. They have been determined by taking different values of the abundance for each element and comparing the results with the disentangled spectra. The nitrogen abundance we found in the primary star is equal to approximately 16 times the solar abundance (taking the solar abundances of \citealt{GAS07}). The nitrogen lines used to obtain this result are \ion{N}{iii}~$\lambda\lambda$~4511,~4515,~4518,~4523,~4528. This nitrogen enhancement is consistent with the overabundance of about 6 obtained with XMM-Newton data for the global X-ray emitting plasma of Plaskett's star \citep{LRP06}. Indeed, in the X-ray data it was not possible to separate the contribution of the primary and secondary plasma in the emission, and thus the overabundance reported by \citet{LRP06} was for the entire system. The lines used to determine the carbon abundance in the primary star are \ion{C}{iii}~$\lambda\lambda$~4070,~4152,~4156,~4163. It appears that there is a clear carbon underabundance (3~\% of the solar abundance) in this component. The N enrichment and C depletion of the primary component are expected from CNO processing in massive stars \citep{MM03}. In the secondary star, it is the helium abundance that shows an overabundance, which is surprisingly coupled with a large underabundance in nitrogen. The abundance in carbon has been fixed to the solar value, because there is no relevant line in the secondary spectrum that could serve for a determination. 
\begin{figure}[t]
  \resizebox{\hsize}{!}{\includegraphics{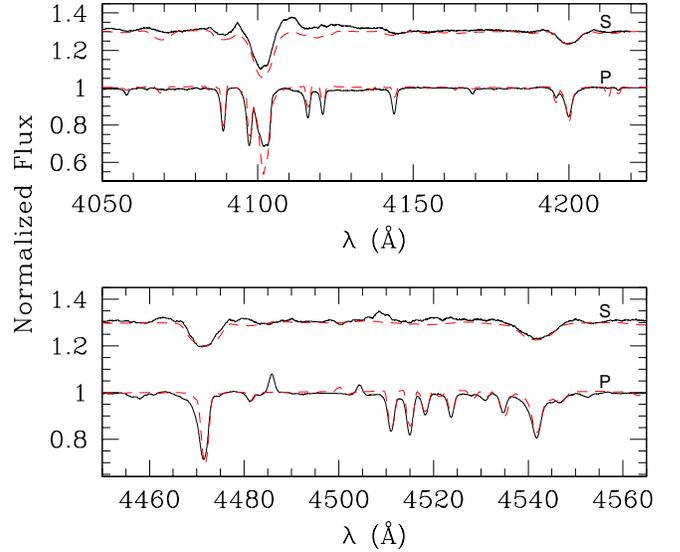}}
  \caption{Comparison between the disentangled optical spectra of \object{HD~47\,129} (solid line) and the theoretical spectra (red dotted line) computed with the CMFGEN atmosphere code. The secondary spectrum was vertically shifted by 0.3 units for clarity.}
  \label{fig:theo_opt}
\end{figure}

\begin{figure}[t]
  \resizebox{\hsize}{!}{\includegraphics{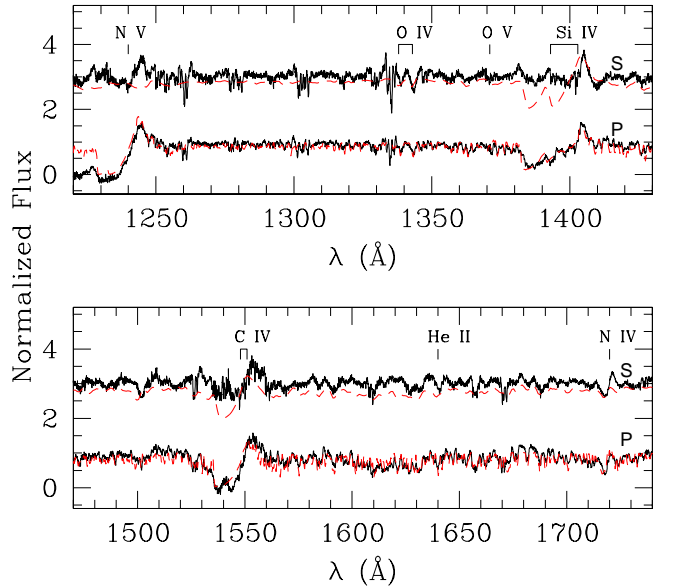}}
  \caption{Same as Fig.~\ref{fig:theo_opt} but for the UV domain. The disentangled spectra have been built from IUE archive spectra of \object{HD~47\,129}. The secondary spectrum was vertically shifted by 1.5 units for clarity.}
  \label{fig:theo_uv}
\end{figure}

Figs.~\ref{fig:theo_opt} and~\ref{fig:theo_uv} show the comparison between the observed and the theoretical spectra in the optical and UV domain, respectively. The correspondance between the optical part of the theoretical spectrum and the disentangled spectra is good for both components. For the IUE spectra, the agreement with the primary component is quite good, but the secondary star spectrum is not correctly represented and the secondary's wind parameters could not be determined, due to the puzzling shape of the UV spectrum of the secondary, where the classical wind lines (\ion{N}{v}~$\lambda$~1240, \ion{Si}{iv}~$\lambda$$\lambda$~1393,~1403, \ion{C}{iv}~$\lambda$~1548,~1551) are not seen. The X-ray luminosity has been taken into account in the models, which allows a better fit of the \ion{N}{v}~$\lambda$~1240 line ($\log\frac{L_X}{L_{bol}} = -7$, \citealt{CG91}) although the blue part of the line is contaminated by the Lyman $\alpha$ interstellar absorption line.

\section{The evolutionary status of \object{HD~47\,129} \label{evolution}}

Some of the most important properties of Plaskett's Star suggest that it is currently in a post Case A RLOF evolutionary stage. The first one is the high rotational velocity of the secondary star. In fact, when the component that fills up its Roche lobe transfers material to its companion, there is also a huge amount of angular momentum that is simultaneously transferred. As a result, the accretor is spun up, whilst the mass donor spins down \citep[e.g.][]{Wellstein2001,PLvdH05}. In a wide binary system, the accretion efficiency can be rather low: only a fraction of the mass lost by the donor is actually accreted, but this is nevertheless sufficient to spin up the accretor to substantial rotational velocities. The accretion efficiency is also reduced by the fast rotation of the accretor. Indeed, accretion would fully stop if the accretor is spun up to its critical velocity \citep{PLvdH05}. In conclusion, the secondary of \object{HD~47\,129} was likely spun up to its present-day high rotational velocity during a previous mass transfer event. The fact that $v\,\sin{i}$ remains high is an indication that either the mass transfer stopped only very recently, or the synchronisation of the rotation due to tidal interaction is very inefficient in this case. If the average accretion efficiency was indeed as low as 10\%, then the present-day properties of \object{HD~47\,129} could only be explained if the primary was initially very massive (around 80\,M$_{\odot}$) and the secondary had about 60\,M$_{\odot}$. However, it must be stressed that a system with as high a primary mass would be expected to undergo an LBV phase rather than a RLOF \citep{Vanbeveren1991}.

Second, the He overabundance of the secondary star is likely related to a previous RLOF event. Indeed, the secondary appears less evolved than the primary in terms of CNO processing: while the latter clearly shows a N overabundance and a C depletion typical of an advanced state of chemical evolution (e.g.~\citealt{MM03}), the former appears rather unevolved, at least in terms of N enrichment. On the other hand, the secondary component has a higher He content compared to the primary for which He/H has not yet changed from its initial value. From this, one can conclude that the He enrichment of the secondary star is not the result of classical evolution, but must originate from a mass transfer. The only caveat in this scenario is the underabundance of nitrogen in the secondary which, at present, cannot be explained by a RLOF event. \citet{HBL08} discuss the surface abundances of a sample of early O-type stars in the LMC as a function of their projected rotational velocities. Whilst most of the objects showed a behavior consistent with the expected chemical enrichment of the surface layers due to rotational mixing, \citet{HBL08} identify two groups of stars with peculiar properties. Their group~1 consists of O-type stars that are fast rotators but do not show a significant nitrogen enhancement in their spectra. On the contrary, group~2 of \citet{HBL08} contains O-type stars that display a rather low $v\,\sin{i}$, whilst simultaneously showing a significant nitrogen enrichment. The properties of these objects appear somewhat related to those of the components of Plaskett's star. The rotational velocity and the surface composition of the primary star of \object{HD~47\,129} are reminiscent of those of the group~2 stars, whereas the secondary component of \object{HD~47\,129} has properties that would place it among group~1 objects. Furthermore, \citet{LCY08} noted that mass transfer can produce rapidly rotating stars that are not strongly nitrogen-enriched if the transfer is highly non-conservative.

\section{Conclusion \label{conclusions}}

We have presented a spectroscopic analysis of the massive O-type binary \object{HD~47\,129}. 

The radial velocities of both components have been inferred using a disentangling method. The orbital solution gave two massive components of 54~M$_{\odot}$ and 56~M$_{\odot}$ for the primary and the secondary star respectively (assuming an inclination of 71\degr). The projected rotational velocities, determined using a Fourier technique, are very different from each other ($\sim$~75~km~s$^{-1}$ for the primary, and~$\sim$~300~km~s$^{-1}$ for the secondary). 

The spectral types, calculated from the two disentangled spectra, gave O8~III/I~+~O7.5~III, which are too late spectral types for the dynamical masses. This discrepancy cannot be solved by taking into account the large rotational velocity of the secondary star. Whilst this effect indeed acts on the secondary luminosity, it is not sufficient to explain such a difference and it explains nothing concerning the primary star.

The optical spectra display a phase-locked variability in the strong emission lines \ion{He}{ii}~$\lambda$~4686 and H$\alpha$. A Doppler tomography technique allowed us to obtain the Doppler maps of both lines which present their highest peak near the primary component velocity. Its velocity coordinates indicate a motion from the secondary toward the primary, which can be interpreted by a colliding wind interaction, where the shock is displaced toward the primary star. The Doppler map also shows an annular feature, almost centered on the system's center of mass, that seems to be another consequence of the secondary's rotation. Indeed, it likely represents the secondary's wind compressed into its equatorial plane by the fast rotation. 

Furthermore, we used the high resolution spectra in order to study the Struve-Sahade effect in this system. Actually, while Plaskett's star presents strong variations in its absorption lines, it is difficult to connect them to the 'standard' definition of the Struve-Sahade effect, which is: ''the apparent strengthening of the secondary spectrum of the binary when it is approaching, and the corresponding weakening of its lines when it is receding'' \citep{BGR99}. The variations are visible in both stars, even though they are stronger in the secondary's spectrum, but it is not clear that they are phase-locked. They appear dominated by a modulation that is either stochastic or at least not tied to the orbital phase. If correct, the interpretation of a flattened secondary wind would imply that these apparent variations result from the effect of the turbulence and other velocity perturbations on the emissions formed in the disk around the secondary star. The deformation of the star itself due to the rotation can also be in this case an additional, but not determining, perturbation. In the case of the main sequence stars of \citet{LRS07}, the Struve-Sahade effect was already found to be due to an inhomogeneous distribution of the line formation, because of the deformation of the star and/or mutual heating, in at least three out of four studied objects (\object{HD~100\,213}, \object{HD~159\,176} and \object{DH~Cep}).

Finally, we compared our observations with theoretical spectra computed with the model atmosphere code CMFGEN. We then determined effective temperature and chemical abundances for both components, and wind properties for the primary star. The nitrogen overabundance in the primary star, already detected in the XMM-Newton data, is confirmed in the optical spectra. This comes with an helium overabundance of the secondary star, which suggests a past RLOF in the system, from the primary star to its companion.

In conclusion, we can say that most of the peculiarities in the spectra of \object{HD~47\,129} can be explained today by the evolutionary status of the system (a post-case A RLOF) and by the fact that the projected rotational velocity of the secondary star is quite important. However, the large dynamical masses of both components are still not compatible with their spectral types and luminosities. 

\begin{acknowledgements}
N. L. acknowledges the financial support of the Belgian "Fonds pour la Formation \`a la Recherche dans l'Industrie et dans l'Agriculture". This research is supported in part through the XMM/INTEGRAL PRODEX contract. Thanks to J.P. Degr\'eve for interesting discussion about the evolutionary status of this system.
\end{acknowledgements}

\bibliographystyle{aa}

\end{document}